\documentclass[pre,reprint,superscriptaddress, onecolumn, 11pt]{revtex4-1}

\usepackage{color,amsthm,amsmath,amsfonts,graphicx,bm,amssymb,hyperref}
\usepackage{epstopdf}
\usepackage{comment}
\usepackage{etoolbox}
\usepackage{youngtab}
\usepackage{verbatim}
\usepackage{MnSymbol}

\graphicspath{{Figures/}}

\newcommand{\beq}{\begin{equation}}
\newcommand{\eeq}{\end{equation}}
\newcommand{\bea}{\begin{eqnarray}}
\newcommand{\eea}{\end{eqnarray}}

\newcommand{\footnotehold}[1]{\footnote{#1}\addtocounter{footnote}{-1}\addtocounter{Hfootnote}{-1}}

\def\pv{\boldsymbol\mu}

\def\xv{\mathbf{x}}

\def\kv{\mathbf{k}}
\def\mv{\mathbf{m}}

\def\zero{\boldsymbol{0}}
\def\bv{\boldsymbol\beta}
\def\ZZ{\mathcal{Z}}

\def\cbar{\bar c}
\def\partN{n}
\def\partM{m}
\def\partNS{n_s}
\def\QQ{\Lambda}

\def\sym{\operatorname{sym}}
\def\asym{\operatorname{asym}}

\def\AA{\mathcal{A}}
\def\aA{A}
\def\hA{A^{(h)}}

\def\pp{p}
\def\dis{\eta}

\newcommand{\titleinfo}{Crossing probability for directed polymers 
in random media: exact tail of the distribution} 
\begin{document}

\title{\titleinfo} 

\author{Andrea De Luca}
\email{andrea.deluca@lptms.u-psud.fr}
\affiliation{
LPTMS, CNRS, Univ. Paris-Sud, Universit\'e Paris-Saclay, 91405 Orsay, France}

\author{Pierre Le Doussal}
\email{ledou@lpt.ens.fr}
\affiliation{Laboratoire de Physique Th\'eorique de l'ENS, CNRS \& Ecole Normale Sup\'erieure de Paris, Paris, France.}

\date{\today}

\begin{abstract}	
We study the probability $p \equiv p_\eta(t)$ that two directed polymers in a given random potential $\eta$ 
and with fixed and nearby endpoints, do not cross until time $t$. This probability is itself a random variable 
(over samples $\eta$) which, as we show, acquires a very broad probability distribution at large time.
In particular the moments of $p$ are found to be dominated by atypical samples where $p$ is of order
unity. Building on a formula established by us in a previous work using nested Bethe Ansatz and Macdonald process methods, 
we obtain analytically the leading large
time behavior of {\it all moments} $\overline{p^m}\simeq \gamma_m/t$. From this,
we extract the exact tail $\sim \rho(p)/t$ of the probability distribution 
of the non-crossing probability at large time. The exact formula is compared to numerical
simulations, with excellent agreement. 
\end{abstract}


\maketitle

\section{Introduction}

\subsection{Overview} 

The problem of directed paths, also called directed polymers, in a random potential arises 
in a variety of fields\cite{huse1985huse, 
*kardar1987scaling, *halpin1995kinetic,blatter1994vortices,lemerle1998domain,SoOr07,bec,gueudre2014explore,hwa1996similarity,otwinowski2014clonal}.
In its continuum version, it is connected to the Kardar-Parisi-Zhang (KPZ) growth equation~\cite{kardar1986dynamic} by an exact mapping, the Cole-Hopf transformation. Recent progress in integrability of the KPZ equation in one dimension 
\cite{spohnKPZEdge,calabrese2010free,dotsenko,corwinDP,borodin2014macdonald,flat,SasamotoStationary,Quastelflat,calabreseSine}
have thus been accompanied by new exact results for the directed polymer (DP)
in $1+1$ dimension. Methods from physics such as replica and the Bethe Ansatz
\cite{kardar1987replica,calabrese2010free,dotsenko,flat,SasamotoStationary}, or from mathematics such as the Macdonald processes 
\cite{borodin2014macdonald}, 
led to many exact results both 
for the KPZ and the DP problem. 
Examples in the later case are distributions of the free energy, of the endpoint position \cite{Endpoint}, as well as 
some multi-point correlations \cite{2point}. 

Despite these progresses, many interesting DP observables still evade exact calculations. 
This is the case for instance of quantities testing the spatial structure of the manifold of 
DP ground states
such as the statistics of coalescence times \cite{Pimentel}, or of their low lying excited states, 
such as the overlap and the droplet probabilities, of
great interest for many applications, e.g. to quantum localization
\cite{markus_magneto}. Similarly, very few results are available 
for the problem of several interacting DP which are mutually competing within the same random 
potential, most notably the case of several DP subjected to the constraint of 
non-crossing \cite{natter,emig2001probability,ferrari2004polynuclear,borodin2014macdonald,Doumerc}. 
More generally, not much is known about crossing or non-crossing probabilities for paths in random media.
Since in a random potential directed polymers compete for the same optimal configuration(s), one
can expect that the non-crossing probability may be small. It remains to quantify how small they are and
how rare are the samples such that they are not small.

In a recent work we introduced a general framework to calculate non-crossing probabilities for directed polymers, equivalently free energies of a collection of directed paths with a non-crossing constraint. Specifically,
we studied the probability $p_\eta(t)$ that two directed polymers in the same white noise random potential $\eta \equiv\sqrt{2 \bar c}  ~  \eta(x,t)$
and with all four endpoints fixed nearby $x=0$ (see below precise definition)
do not intersect up to time $t$. We used
the replica method to map the problem onto the 
Lieb-Liniger model with attractive interaction $c = -\cbar < 0$ and generalized statistics between particles.
Employing both the Nested Bethe Ansatz and known formula
from Macdonald processes, we obtained a general formula 
for the integer moments $\overline{p_\eta(t)^m}$ (overbar denotes averages with respect to $\eta$)
which we could relate, at least at a
formal level, to a Fredholm determinant. 
While explicit evaluation of this formula for any $m,t$ appeared very difficult, we were
able to obtain explicit results for $m=1,2$ for all time $t$, and for $m=3$ in the large time limit.
This led us to conjecture that, at large time:
\bea \label{conj1} 
\overline{p_\eta(t)^m} \simeq_{t \to \infty}  \gamma_m \frac{\cbar^{2(m-1)}}{t} 
\eea 
where $\bar c$ is the strength of the disorder, with explicit values for the first three coefficients 
\bea \label{lowmom} 
\gamma_1=\frac{1}{2} \; , \quad \gamma_2=\frac{1}{12} \; , \quad \gamma_3=\frac{1}{15} \;.
\eea 
The calculation of all the $\gamma_m$ and the more general question of the determination of the full probability distribution, ${\cal P}_t(p)$ of
$p \equiv p_\eta(t)$, remained open problems. An interesting finding of \cite{de2015crossing} is that the first moment is {\it exactly} given by
(\ref{conj1}) i.e. $\overline{p_\eta(t)} = 1/(2 t)$ for all $t$, independent of
the disorder strength, and in fact identical to the result without disorder. As explained there (and recalled
below) this arises as a consequence of an exact symmetry of the problem, 
called the statistical 
tilt symmetry (STS). 

\subsection{Aim and main results} 

\begin{figure}
  \includegraphics[width=.9\textwidth]{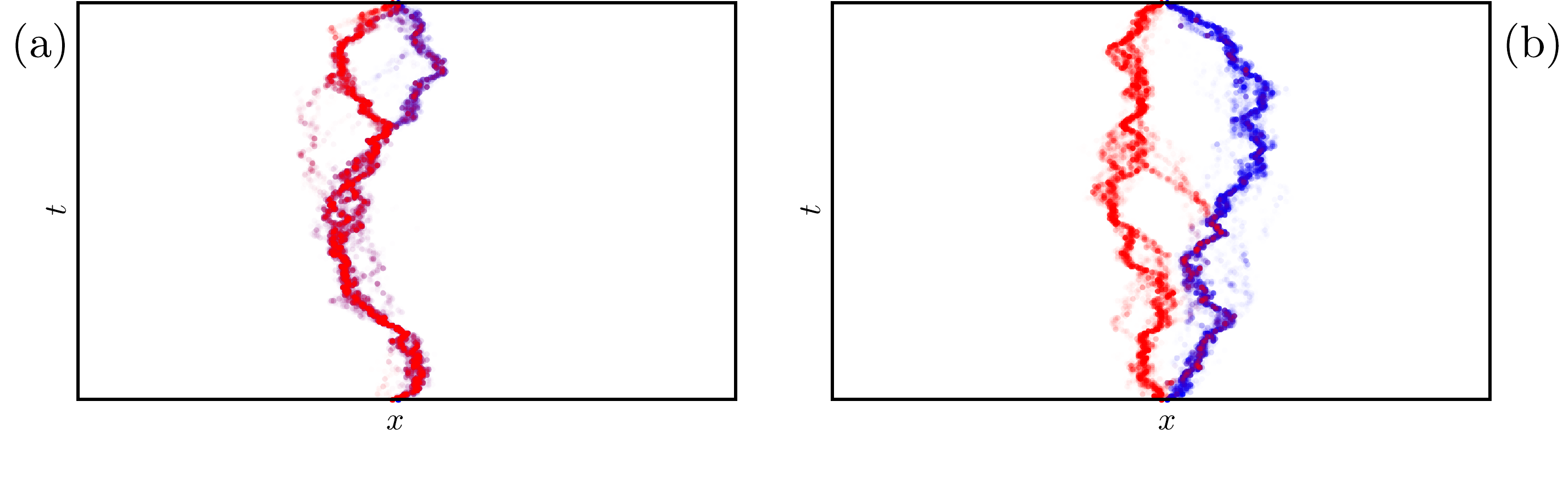}
\caption{
\label{fig1}
Configuration of two polymers starting at $(\hat x=\mp 1/2, \hat t = 0) $ (respectively red/blue) for
a typical value of $p$ in (a) and for a rare realization with large
$p = O(1)$ in (b). The opacity of each dot in $(\hat x,\hat t)$ corresponds
to $Z(\pm 1/2, \hat x | \hat t) \times Z(\hat x, \pm 1/2| \hat \tau - \hat t)$ with $\hat \tau = 300$.
See Section \ref{sec_comparison} for definitions of notations on the lattice.
Here, $\beta = 1.0$ and the red points have been drawn on top of the blues, which 
explains the apparent asymmetry in (a). 
}
\end{figure}

The aim of this paper is to report a first step in the determination of 
the  sample to sample distribution of non-crossing probability
${\cal P}_t(p)$. We will start from the general formula for the moments derived in
\cite{de2015crossing} in terms of multiple integrals over so-called string rapidities, $\mu_j$, of
a quite complicated symmetric polynomial of these rapidities (called $\Lambda_{n,m}(\mu)$ below). 
We will develop general algebraic methods to deal with these types of polynomials and
integrals, and apply them here to study the replica limit $n=0$ and the large time limit. 
We demonstrate that the conjecture (\ref{conj1}) is indeed correct and obtain all the coefficients
$\gamma_m$. From the moments (\ref{conj1})  we are able reconstruct 
an interesting and non-trivial information about the probability distribution ${\cal P}_t(p)$, namely its tail, as we now explain. 

\begin{figure}[ht]
  \includegraphics[width=.5\textwidth]{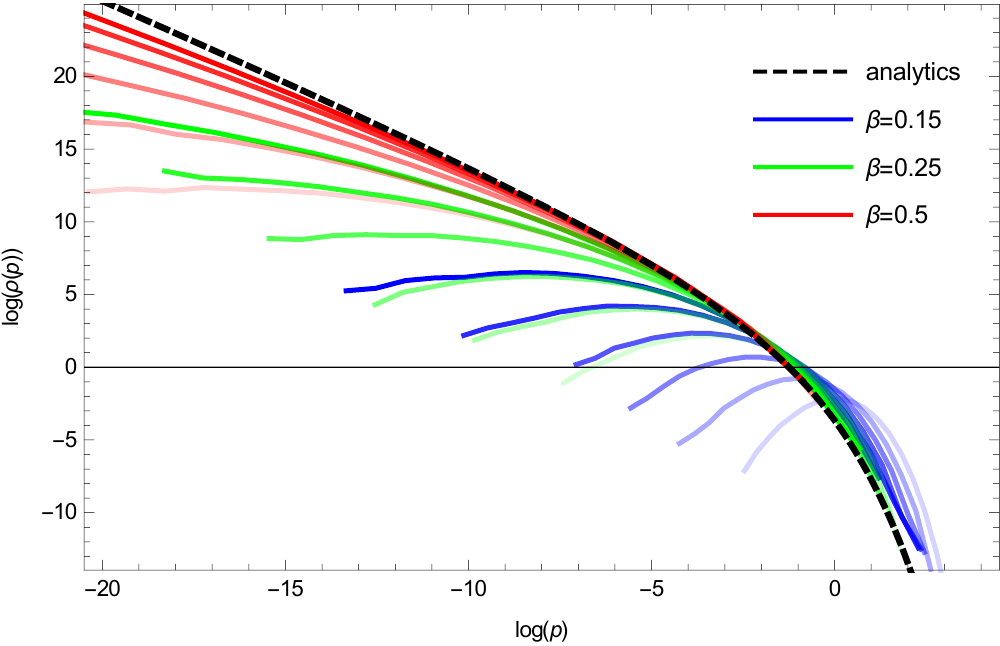}
\caption{
\label{comparison}
Comparison of our prediction for the continuum model 
with numerical simulations of the DP on the square lattice, 
as described in the text. We use at least $2\times 10^5$ realizations of the disorder (samples).
The $\log$ of the empirical distributions of $p = \hat{p}/(16\beta^4)$,
with $\hat{p}$ the probability on the lattice, is shown for different values of $\beta$ and several times
$\hat t = 2^8, 2^9, \ldots, 2^{13}$. The opacity of each line is scaled proportionally $\hat t$. 
The analytical prediction in Eq.~\eqref{rho1} for the tail of the distribution is confirmed at large times $\hat t\to \infty$ and $\beta \to 0$.
Note that the bulk of the distribution and $\ln p_{typ}(t)$ shifts very rapidly to large negative 
values (which we find consistent with $t^{1/3}$, but not analyzed here).
}
\end{figure}

It is important to point out that the result (\ref{conj1}) is valid only for fixed integer $m$ in the limit of
large time. In fact, this knowledge of the leading behavior of the integer moments at large time 
is not sufficient to reconstruct the full distribution of $p$. As we have argued in \cite{supplmat_deluca2015} on the basis of
universality from the results of \cite{Doumerc}, we expect that 
\bea
\ln p_{typ}(t) \equiv \overline{\ln p_\eta(t)} \sim - a  (\cbar^2 t)^{1/3} 
\eea 
where, furthermore,
$a=\overline{\chi_2} - \overline{\chi'_2} \approx 1.9043$ is the average gap between the first ($\chi_2$)
and second ($\chi'_2$) GUE largest (properly scaled) eigenvalues of a random matrix belonging to the
Gaussian unitary ensemble (GUE).
This means that in a {\it typical} realization of the random potential $\eta$,
$p\equiv p_\eta(t)$ is sub-exponentially small at large time, i.e. $p_{typ}(t) \sim e^{- a  (\cbar^2 t)^{1/3} }$.
To account for the form (\ref{conj1}) the integer moments 
should be dominated by a small fraction $\sim 1/(\cbar^2 t)$ of 
environments for which typically $p_\eta(t) \sim \cbar^2$. Hence we are led to 
conclude that
\bea \label{conj2} 
{\cal P}_t(p) \simeq_{t \to +\infty} {\cal P}^{0}\bigl(p/p_{typ}(t)\bigr) + \frac{\rho(p/\cbar^2)}{\cbar^4 t} 
\eea 
where $\rho(p/\cbar)$ is a fixed function and ${\cal P}^{0}$ is the bulk of the
distribution centered around the typical value. Here our goal is to calculate only the tail function $\rho(p)$,
leaving the determination of the bulk function to future studies. 
We obtain, from an exact calculation of the $\gamma_m$ (given in formula \eqref{finalresult} below), 
\bea \label{conj3} 
&& \rho(p) = \frac{2}{p} \int_0^{+\infty} \frac{du}{\sqrt{ u (u+4)}} K_0(2 \sqrt{p} \sqrt{u+ 4})
\eea 
where $K_0$ is the modified Bessel function. It is easy to see that
this result reproduces $\int dp p^m \rho(p) = \gamma_m$ in agreement with 
the values in Eq.~(\ref{lowmom}). 
The conjecture (\ref{conj2}) with the analytical form (\ref{conj3}) is fully confirmed
by our numerical study, see Sec. \ref{sec_comparison}; 
in particular Fig.~\ref{comparison}
shows comparison with the model defined on the square lattice, which at high temperature
is a good approximation of the continuum one.
\\

Strictly Eq.~(\ref{conj2}) and (\ref{conj3}) are 
valid only at fixed $p$ for large $t$, and the total weight in the tail is 
naively $\sim 1/t$. However one sees that the asymptotic behavior 
of the density function $\rho(p)$ at small 
$p$ is
\bea \label{asympt} 
\rho(p) \simeq \frac{1}{2 p} (\ln p)^2 
\eea 
hence its total weight is not integrable at small $p$. Thus
we can surmise that the above form holds for $p > p_c(t)$ 
where $p_c(t)$ is a small-$p$ time-dependent cutoff, and we
can try to match the tail to the bulk around $p_c(t)$. Integration of (\ref{asympt}) 
gives a total weight $\sim \frac{1}{6} |\ln p_c(t)|^3/t $ for the tail region of
the probability distribution. This suggests, assuming no other intermediate scale, the following 
bound on $p_c(t)$: $\frac{1}{6} |\ln p_c(t)|^3/t \ll 1$
i.e. $\ln p_c(t) \gg -(6t)^{1/3} \simeq - 1.817 t^{1/3}$. This is consistent
with $p_{typ}(t) \ll p_c(t)$ but on the same $t^{1/3}$ scale.
A more detailed analysis of this matching is left for the future. \\

Finally one may wonder how the samples with values of $p$ of order one differ in real space from
the ones with typical values of $p$. For this, we show in Fig \ref{fig1} density plots of the configurational
probabilities of two independent directed polymers in the same environment 
constrained to start and end at different, but very close-by
points (nearest neighbors on the lattice). We show two samples: for the sample 
with higher $p$, the small difference in starting points results in a very large 
difference in most probable configurations. The details of the numerics are
discussed in section Sec. \ref{sec_comparison}.
\\

The paper is organized as follows: in Sec.~\ref{sec_model}, we recall
the model, the observables and the main results of \cite{de2015crossing}
which are the starting point for the present calculation; in Sec. \ref{sec_building},
we study the building blocks for the formula of the moments of $p_\eta(t)$;
finally in Sec. \ref{sec_moments}, we apply these formulas in the limit $n\to0$ and 
of large times to derive the coefficients $\gamma_m$ and the distribution of $p_\eta(t)$, which 
is then compared to numerics.

\section{Model, observables and starting formula \label{sec_model}}

\subsection{Model and observables} 

The model of a directed polymer in the continuum in dimension $1+1$ is defined by the partition sum
of all paths $x(\tau) \in \mathbb{R}$ starting from $x$ at time $\tau=0$, and ending at $y$ at time $\tau=t$. 
This can be seen as 
the canonical partition function of a directed polymer of length $t$ with fixed endpoints
\begin{equation}
 \label{Zorig}
 Z_\dis(x;y|t) \equiv \int_{x(0) = x}^{x(t) = y} Dx e^{-\int_0^t d\tau \left[ \frac{1}{4}(\frac{dx}{d\tau})^2 - \sqrt{2 \cbar} \dis(x(\tau), \tau)\right]} 
 \end{equation}
 in a random potential with white-noise correlations $\overline{\dis(x,t) \dis(x',t')} = \delta(x-x')\delta(t-t')$.  
 It describes the thermal fluctuations of a single polymer in a given realization $\eta$ of the random potential (a sample). \\
 
 Thanks to the Karlin-McGregor formula for non-crossing paths and its generalizations
 \cite{karlin1959coincidence, *gessel1985binomial, *gessel1989determinants, *wiki:lgvlem},
the partition sum of two polymers with ordered and fixed endpoints, $(x_1,y_1)$ and $(x_2,y_2)$,
is given by a determinant formed with the single polymer partition sums:
\bea
Z^{(2)}_\eta(x_1,x_2;y_1,y_2|t) = Z_\dis(x_1;y_1|t)Z_\dis(x_2;y_2|t) - Z_\dis(x_2;y_1|t)Z_\dis(x_1;y_2|t) \;.
\eea 
Hence one can express the probability (over thermal fluctuations) that two polymers with fixed endpoints
do not cross in a given realization $\eta$ as the ratio:
\begin{equation}
 \label{pnoncross}
 p_\dis(x_1,x_2;y_1,y_2|t) \equiv 1 - \frac{Z_\dis(x_2;y_1|t)Z_\dis(x_1;y_2|t)}{Z_\dis(x_1;y_1|t)Z_\dis(x_2;y_2|t)} \;.
 \end{equation}
\noindent 
Here for simplicity, we study only the random variable defined by the limit of near-coinciding endpoints
\begin{equation}
 \label{plimit}
 \pp_\dis(t) \equiv \lim_{\epsilon \to 0} \frac{p_\dis(-\epsilon, \epsilon; -\epsilon, \epsilon|t)}{4\epsilon^2}
\end{equation}
As noticed in \cite{de2015crossing}, it can also be written as:
 \label{plimit2}
 \begin{equation}
 \pp_\dis(t) 
 = \left.\partial_x \partial_y \ln Z_{\dis}(x; y|t)\right|_{\substack{x=0\\y=0}}
\end{equation}
which is useful in some cases, e.g. to show that the first moment is independent of
the disorder, see \cite{de2015crossing}. \\

\subsection{Replica trick and starting formula} 

The observables that we will study here are the integer moments of this probability $p_\eta(t)$. Using the replica trick these
moments can be written as
\bea
\overline{\pp_\dis(t)^\partM} = \lim_{n \to 0} \Theta_{n,\partM}(t)
\eea 
where we have introduced:
\begin{equation}
 \label{plimitGen}
 \Theta_{\partN,\partM}(t) \equiv \lim_{\epsilon\to0} \overline{[(2\epsilon)^{-2} Z_\dis^{(2)}(\epsilon)]^\partM [Z_\dis(0;0|t)]^{\partN-2\partM}}
 \end{equation}
and we defined the partition sum of two non-crossing polymers with endpoints near 
$x=0$ as 
\begin{equation}
 \label{Z2}
Z^{(2)}_\dis(\epsilon) \equiv Z_\dis(\epsilon;\epsilon|t)Z_\dis(-\epsilon;-\epsilon|t)-
Z_\dis(-\epsilon;\epsilon|t)Z_\dis(\epsilon;-\epsilon|t)
 \end{equation}

The idea is now to calculate 
$\Theta_{n,\partM}(t)$ and then to take the limit $n=0$.  \\

In Ref. \cite{de2015crossing}, we have derived a formula for these quantities. 
This result was obtained in the simplest case ($m=1$)
by use of the nested Bethe Ansatz and, for general $m$, using a contour integral formula obtained from the
theory of Macdonald processes in \cite{borodin2014macdonald}, with perfect agreement between the two methods. 

The formula goes as follows. For each $n,m$, one first defines a function of a set of $n$ complex variables $\mu_\alpha$, $\alpha=1,..n$, 
the ``rapidities'' (also indicated collectively by a vector $\pv = (\mu_1,\ldots, \mu_\partN)$) as
\begin{equation}
 \label{borodinG}
  \QQ_{\partN,\partM}(\pv) = \frac{1}{2^\partM}\sym_{\pv} 
  \left[\frac{\prod_{q=1}^\partM h(\mu_{2q-1,2q})}{\prod_{1\leq \alpha<\beta\leq \partN} f(\mu_{\beta\alpha})}
  \right]
 \end{equation}
where $\mu_{\beta \alpha} = \mu_{\beta} - \mu_{\alpha}$ and we have introduced the two functions 
\bea \label{hf} 
h(u) = u (u - i c) \quad , \quad f(u) = u/(u-i c)
\eea 
and the symmetrization operator over the variables $\mu$:
 \begin{equation}
  \sym_{\pv} [ F(\mu_1,\ldots,\mu_\partN)] = \frac{1}{n!}\sum_{P\in \mathcal{S}_\partN} F(\mu_{P_1},\ldots,\mu_{P_\partN}) \;.
 \end{equation}
As discussed below,
the rational function in Eq.~\eqref{borodinG} is actually a symmetric polynomial in the $\mu_\alpha$. 
 
The formula obtained in \cite{de2015crossing} then reads:
\begin{equation}
\label{start}
\Theta_{\partN,m}(t) =  \langle \QQ_{\partN, \partM}(\pv) \rangle_{\partN}
\end{equation}
where we introduced the ``string average'' for any symmetric function $F(\pv)$ as
\begin{equation}
\label{stringave}
\langle F(\pv) \rangle_{\partN} \equiv \sum_{\partNS=1}^\partN \frac{\partN! \cbar^\partN}{\partNS! (2 \pi \cbar)^{\partNS}} \sum_{(m_1,..m_{\partNS})_{\partN}} \prod_{j=1}^{\partNS} \int_{-\infty}^{+\infty} \frac{dk_j} {m_j} 
\Phi(\kv, \mv) F^s(\kv, \mv) e^{-\AA_2^s(\kv, \mv) t}  \;.
\end{equation}
In this equation, we introduce the notation
\bea
&&F^s(\kv, \mv)  \equiv  F(\pv)|_{{\bf \mu} = {\bf \mu^s}} \;, \qquad \Phi(\kv, \mv) = \prod_{1\leq j<j'\leq \partNS} \frac{(k_j - k_{j'})^2 + \cbar^2(m_j - m_{j'})^2/4}{(k_j - k_{j'})^2 + \cbar^2(m_j + m_{j'})^2/4} \;,\\
&& \mu^s_\alpha \equiv \mu^s_{j,a} = k_j + \frac{i \cbar}{2} (m_j + 1 - 2a)  \; , \quad a_j = 1,..m_j  \;, \quad  j=1,..n_s \label{string} \;,\\
&& \AA_p(\pv) \equiv \sum_{\alpha=1}^\partN \mu_\alpha^p \; , \quad {\cal A}_p^s(\kv, \mv) 
\equiv  \AA_p(\pv)|_{{\bf \mu} = {\bf \mu^s}}  \label{charges}
\eea
where $A_2$ denotes the energy and $\AA_p$ corresponds to the
conserved charges of the Lieb-Liniger model \cite{ll}.
The factor $F^s(\kv, \mv)$ is obtained from $F(\pv)$ replacing the values 
of the $n$ rapidities $\mu_\alpha$ with their values 
$\mu^s_{j,a}$ for a ``string state'' and so is 
$\AA_p^s(\kv, \mv)$ obtained from $\AA_p(\pv)$. Such a ``string state'' is characterized by: 
(i) an integer $n_s$, the number of strings in the state, with $1 \leq n_s \leq n$;
(ii) $n_s$ real variables $k_j \in \mathbb{R}$, $j=1,\ldots, n_s$, the ``momenta of the string center of mass'';
(iii) $n_s$ integer variables $1 \leq m_j$, "the particle content" of each string in the string state. 
In the above formula (\ref{stringave}) a ``summation'' over all string states is performed, meaning
that these variables are summed upon or integrated upon.
Here, $(m_1, \ldots ,m_{n_s})_n$ indicates sum over all 
integers $m_j \geq 1$ whose sum equals $n$, i.e. $\sum_{j=1}^{n_s} m_j = \partN$. 

An important property of \eqref{start} and \eqref{stringave} is that 
considering $\langle \QQ_{\partN,0} \rangle_{\partN} \equiv \langle 1 \rangle_{\partN} $, 
one recovers the formula for 
$\ZZ_{\partN}(t)\equiv \ZZ_{\partN}(\xv = \zero;\zero| t) = \Theta_{n,0}(t)$ 
for the $n$-th moment of a single DP partition sum with fixed endpoint,
studied and calculated in \cite{calabrese2010free}. The present calculation
is thus a, quite non-trivial, generalization of that calculation. 

The formula \eqref{start} is thus our starting point. 
We now turn to explicit calculations of the building blocks in Eq.~\eqref{borodinG}.

\section{Calculation of the building blocks $\Lambda_{n,m}({\bf \mu})$ \label{sec_building}}

In this section we provide an explicit formula 
for $\QQ_{\partN, \partM}(\pv)$ as a symmetric polynomial.
This approach is based on: (i) the invariance of \eqref{borodinG} under the simultaneous translation of 
all the rapidities $\mu_\alpha \to \mu_\alpha + u$; (ii) the fact that
$\QQ_{\partN,\partM}$ vanishes on any $\ell$-string with $\ell > \partN- \partM$.

The best way to deal with this problem, 
is to separate these polynomials into homogeneous components, which are discovered to coincide with 
the $\QQ_{\partN, \partM}(\pv)$ computed at $\cbar =0$. Hence, we start by studying this case. 

\subsection{$\cbar=0$ case \label{czerocase}}
We define $\tilde{\QQ}_{\partN,\partM} (\pv)$ as $\QQ_{\partN,\partM}(\pv)$ computed at $\cbar=0$. In this case $f(u) = 1$ in (\ref{hf}) and 
therefore Eq.~\eqref{borodinG} simplifies to
\begin{equation}
 \label{freeBoro}
  \tilde{\QQ}_{\partN,\partM}(\pv) = \frac{1}{2^\partM}\sym_{\pv} \left[\prod_{q=1}^\partM (\mu_{2q-1} - \mu_{2q})^2 \right]
  = \sym_{\pv} \left[\prod_{q=1}^\partM (\mu_{2q-1}^2 - \mu_{2q-1}\mu_{2q}) \right]
 \end{equation}
where the second equality is obtained expanding the square and replacing $\mu_{2q} \to \mu_{2q-1}$ inside the symmetrization.
We want to re-express \eqref{freeBoro} in terms of the elementary symmetric polynomials 
\begin{equation}
 \label{esymmdef}
 e_{p}(\pv) = \sum_{1 \leq \alpha_1 < \ldots < \alpha_p \leq \partN } \mu_{\alpha_1} \ldots \mu_{\alpha_p} \;.
\end{equation}
with $e_0(\pv)=1$ and we will use below the convention that $\mu_\alpha=0$ for $\alpha>n$, leading to 
$e_{p}(\pv)=0$ for $p>\partN$	. We will also omit the explicit dependence on rapidities when these do not take 
a specific value and simply denote $e_p \equiv e_{p}(\pv)$. 
It is important to underline few properties that \eqref{freeBoro} has to satisfy. Indeed,
$\tilde{\QQ}_{\partN, \partM}(\pv)$ is a polynomial 
\begin{enumerate}
 \item symmetric in the variables $\mu_1,\ldots, \mu_\partN$;
 \item homogeneous of degree $2\partM$;
 \item containing each rapidity $\mu_\alpha$ with degree at most $2$;
 \item invariant under a simultaneous translation of all variables: $\tilde{\QQ}_{\partN,\partM}(\pv + u) =  \tilde{\QQ}_{\partN,\partM}(\pv)$
 for any complex number $u$ and $\pv + u = (\mu_1 + u, \ldots, \mu_\partN + u)$.
\end{enumerate}
Conditions 1, 2, 3 impose that $\tilde{\QQ}_{\partN, \partM}(\pv)$
is a linear combinations of terms $e_{p} e_{2m-p}$ with $p=0,\ldots,m$. 
Moreover, in this expansion, all the coefficients, but one, can be fixed
using condition 4. An important consequence, which we will use below, is that, for any given $\partN,\partM$, 
a polynomial satisfying conditions 1--4 has to be
a multiple of $\tilde{\Lambda}_{\partN, \partM}$.
Additionally, by focusing on the coefficient of $\prod_q \mu_{2q-1}^2$
in \eqref{freeBoro}, it can be seen that $e_{m} e_m$ appears multiplied by $m! (n-m)!/n!$. 
We refer to Appendix \ref{app_tildel} for all the details
and we get
\begin{equation}
\label{LambdatildeFin}
   \tilde{\QQ}_{\partN,\partM}(\pv) = \frac{m!}{n! (n-m)!} (-1)^m \sum_{p=0}^{2m} (-1)^p (n-p)! (n-2m+p)! e_{p} e_{2m-p} 
\end{equation}
which is the required expansion. 
Remarkably, this expression is a convolution and can therefore be expressed compactly using generating functions. We 
recall the standard generating function for the elementary symmetric polynomials
\begin{equation}
 E(x | \pv) = \prod_{i\geq 1} (1 + \mu_i x) = \sum_{r\geq 0 } e_r x^r \;.
\end{equation}
Again we will write simply $E(x | \pv) \equiv E(x)$, and similarly for other generating functions, 
whenever the rapidities are
considered at generic values. 
Then, we can rewrite
\begin{equation}
\label{LambdatildeGen}
   \tilde{\QQ}_{\partN,\partM}(\pv) = \frac{m!}{n! (n-m)!} (-1)^m [H_{\partN}(x)H_{\partN}(-x)]_{x^{2\partM}}
\end{equation}
where we introduced
\begin{equation}
 \label{Hfun}
 H_\partN(x) = \sum_{p=0}^n (\partN-p)! e_p x^p = \int_0^\infty dt e^{-t} t^n E(x/t)
\end{equation}
and everywhere here $[F(x)]_{x^p}$ indicates the coefficient of $x^p$ in the series $F(x)$.

\subsection{General case $\bar c \neq 0$}
The study of the finite $\cbar$ case requires a more detailed analysis. First of all, 
one can check (see Appendix \ref{app_residue})
that $\Lambda_{n,m}(\pv)$ is still a symmetric polynomial in the rapidities. Moreover it is homogeneous 
of degree $2m$ in the combined set of $\cbar, \mu_1,\ldots, \mu_{\partN}$. Thanks to the characterization of 
$\tilde{\Lambda}_{n,m'}(\pv)$ in terms of properties 1--4 given in the previous section, it can be seen 
(Appendix \ref{app_taylorc}) 
that $\Lambda_{\partN, \partM}(\pv)$ admits the following expansion 
 \begin{equation}
\label{recurstilde}
 \QQ_{\partN, \partM} (\pv) = 
 \sum_{a=0}^{\partM} \cbar^{2a} \Omega_{\partN,\partM}^a \tilde{\QQ}_{\partN, \partM-a} (\pv) \;,
\end{equation}
where the $\Omega_{\partN,\partM}^a$ are constant coefficients, for the moment unknown,
expect for $\Omega_{\partN, \partM}^0 = 1$. 
Thanks to Eq.~\eqref{LambdatildeGen}, it is possible to rewrite 
Eq.~\eqref{recurstilde} again in terms of generating functions as
\begin{equation}
\label{recursGen}
\QQ_{\partN, \partM} (\pv) = 
\frac{m!(-1)^m}{(n-m)!n!} [\omega_{n,m}(i \cbar x) H_n(x) H_n(-x)]_{x^{2m}}
\end{equation}
where we introduced the generating function of the unknowns $\Omega_{n,m}^a$
\begin{equation}
\label{omegagen}
 \omega_{n,m}(x) = \frac{(n-m)!}{m!}\sum_{a=0}^m \Omega_{\partN,\partM}^a 
\frac{(m-a)!}{(n-m+a)!} x^{2a}
\end{equation}
with $\omega_{n,m}(0) = 1$. 
Since Eq.~\eqref{recurstilde} and Eq.~\eqref{recursGen} holds
for an arbitrary choice of $\pv$,
the values of the $\Omega_{\partN, \partM}^a$ can be fixed 
by choosing specific configuration of 
rapidities where $\Lambda_{\partN, \partM}(\pv)$ simplifies. 
Consider in particular the string configuration (\ref{string}) 
characterized by $m_1 = \ell$ and $m_2 = \ldots = m_{n - \ell +1}=1$, all with vanishing momenta $k_j=0$, i.e.
\begin{equation}
\label{stringlzero}
\pv^{\ell,0} = \left( \frac{i \cbar }{2} (\ell - 1), \frac{i \cbar }{2} (\ell - 3), \ldots,  -\frac{i \cbar }{2} (\ell - 1), 0, \ldots, 0\right) \;,
\end{equation}
then one obtains (see Appendix \ref{app_strings}) that
 \begin{equation}
\label{substl}
 \QQ_{\partN, \partM} (\pv^{\ell,0} ) = 0\;, \qquad \mbox{for any} \; \ell = n-m+1,n-m+2, \ldots,n \;.
\end{equation}
These conditions have a direct physical interpretation: in the Lieb-Liniger 
language, an $\ell$-strings can be considered
as a bound-state composed by $\ell$ particles; in order to form a string with $\ell > n-m$, necessarily, 
the rapidities of two-particles which are mutually avoiding each other would need to be included in the string.
As no bound state can be formed between avoiding particles, this term gives a vanishing contribution in Eq.~\eqref{start}. 
So, the condition expressed by \eqref{substl} encodes the effective repulsion 
between polymers. The value of the elementary symmetric polynomials for this configuration $\pv =\pv^{\ell,0}$ can be
found explicitly (see Appendix \ref{app_omega}) as
\begin{equation}
 \label{symstring}
 e_p (\pv^{\ell,0}) = \binom{\ell}{p}(- i \cbar)^p B_p^{(\ell+1)}\bigl(\frac{\ell+1}{2}\bigr)\;.
\end{equation}
We will extensively use in this paper the {\it generalized Bernoulli polynomials} 
\footnotehold{The $B_n^{(\alpha)}(y)$ are indicated as 
NorlundB[$n,\alpha,y$] in \textit{Mathematica} .}
which have been introduced from the generating function
\begin{equation}
 \label{genbern}
 G_\alpha(x,y) \equiv \left(\frac{x}{e^x -1}\right)^\alpha e^{xy}  = \sum_{n=0}^\infty \frac{B_n^{(\alpha)}(y) x^n}{n!} \;.
 \end{equation}
By inserting Eq.~\eqref{symstring} in Eq.~\eqref{Hfun} and 
denoting $H_{\partN}^{(\ell)}(x) = H_\partN(x | \pv^{\ell,0})$,
we arrive at (see Appendix \ref{app_omega})
\begin{equation}
\label{HH}
H^{(n-k)}_{\partN}(x)H^{(n-k)}_\partN(-x) = \sum_{p=1}^{2k+1} b_p G_{2n+2}(i \cbar x, n-k+p)
\end{equation}
where $k = \partN - \ell$ and the coefficients $b_k$ satisfy the symmetry $b_{2k + 2 - p} = b_p$
as is seen from the property 
\bea \label{Gsym}
G_\alpha(x,y)=G_\alpha(-x,\alpha-y)
\eea
and the
fact that the left-hand side of \eqref{HH} is an even function of $x$. Using Eq.~\eqref{HH}
and Eq.~\eqref{recursGen}, the conditions in Eq.~\eqref{substl} are equivalent to
\begin{equation} \label{newconditions} 
[ \omega_{n,m}(icx) (G_{2n+2}(i \cbar x, n-k +1) + G_{2n+2}(-i \cbar x, n-k +1)) ]_{x^{2m}} =
 0\;, \qquad \forall \; k = 0, \ldots, m-1
\end{equation}
To see this, we start from $k=0$, in which case there is only one
term in the sum (\ref{HH}). Then, for $k=1$, the sum involves three terms. However, 
using the condition for $k=0$, we can reduce to the first and last term
in the sum and 
obtain (\ref{newconditions}),
again by the symmetry in Eq.~(\ref{Gsym}). Similarly, one can proceed for all $k$ up to $k=m-1$ using each time, all the previous 
conditions up to $k-1$.

These conditions (\ref{newconditions}) are solved 
by 
\bea \label{soluomega} 
\omega_{n,m}(x) = \frac{1}{2} (G_{2m-2n-1}(x, m-n) + G_{2m-2n-1}(-x,m-n))  + O(x^{2m+2})
\eea 
where the higher orders do not affect
the $x^{2m}$ 
coefficients needed in \eqref{recursGen}.
To see that Eq.~\eqref{soluomega} satisfies \eqref{newconditions}, we use that
\bea
B_{2m} ^{(2m+1)}(m-k + 1)= (-k-m + 1)_{2 m} = 0 \;, \qquad \forall k = 0,\ldots, m-1 
\eea 
where $(x)_p=x(x+1)..(x+p)$ is the Pochhammer symbol, as shown in detail in the 
Appendix \ref{app_genbern}. Finally, we get, 
from Eq.~\eqref{soluomega}  and Eq.~\eqref{genbern}, our final explicit expression for the coefficients $\Omega_{n,m}^a$ as
\begin{equation}
\label{omegasol}
\Omega_{n,m}^a = \frac{m! (n-m+a)! B_{2 a}^{(2 m-2 n-1)}(m-n)}{(2 a)! (m-a)! (n-m)!} \;.
\end{equation}
in terms of generalized Bernoulli polynomials,
which complete the expansion of $\QQ_{\partN, \partM}(\pv)$ in Eq.~\eqref{recurstilde}. 
More compactly we can write 
combining \eqref{recursGen} and \eqref{soluomega}
\begin{equation}
 \label{lambdacompact}
\QQ_{\partN, \partM} (\pv) = 
\frac{m!(-1)^m}{(n-m)!n!} \left[
\left(\frac{\cbar x}{2 \sin \frac{\cbar x}{2}}\right)^{2 m-2 n} \frac{\sin \cbar x}{\cbar x} 
H_n(x) H_n(-x)\right]_{x^{2m}} 
\end{equation}
which is the main result of this section.

\section{Calculation of the moments of $p$ \label{sec_moments}}
\subsection{$n\to0$ limit}
Thanks to the results of the previous section, we can now express $\QQ_{\partN, \partM}(\pv)$ 
in terms of the elementary symmetric polynomials. Then, the 
dependence in terms of the
conserved charges $\AA_p$ in Eq.~\eqref{charges}
can be recovered using the Newton's identities \cite{macdonald1995symmetric}
\begin{equation}
\label{newt}
 e_p = \frac1{p!} \det \begin{pmatrix}
    \AA_1     & 1       & 0      & \cdots       \\
    \AA_2     & \AA_1     & 2      & 0      & \cdots \\
    \vdots  &         & \ddots & \ddots       \\
    \AA_{p-1} & \AA_{p-2} & \cdots & \AA_1    & p-1 \\
    \AA_p     & \AA_{p-1} & \cdots & \AA_2    & \AA_1
  \end{pmatrix} 
\end{equation}
Therefore, as explained in \cite{de2015crossing}, after introducing the generalized replica partition function
\begin{equation}
\label{GGEpartfun}
\ZZ_{\partN}^{\bv}(t) 
=  \langle e^{\sum_{p \geq 1} \beta_p \AA_p} \rangle_{\partN} \;,
\end{equation}
the relation for $\Theta_{\partN, \partM}$ in 
Eq.~\eqref{start} can be rewritten as
\begin{equation} 
\label{sub} 
 \Theta_{\partN,m}(t)= \QQ_{\partN,m}[\{\partial_{\beta_p}\}] [ \ZZ_\partN^{\bv} (t) ]  \;.
\end{equation}
Here, we first define $\Lambda_{n,m}[\AA_p]$ as $\Lambda_{n,m}(\pv)$ expanded as a function
of the $\AA_p$, for simplicity without using a new symbol.
Then, we formally replace in $\QQ_{\partN,\partM}[\AA_p]$ the charges $\AA_p \to \partial_{\beta_p}$, with
the derivatives computed 
setting all $\beta_p$'s to zero afterwards.
In the limit $\partN\to 0$ prescribed by the replica trick, we can write
\begin{equation}
\QQ_{\partN,\partM}(\pv)  =  \frac{\lambda_{\partM}(\pv)}{n} + O(\partN^0) \;.
\end{equation}
and neglect all the subleading orders in the Taylor expansion in powers of $\partN$,
as they act as derivatives of a constant $\lim_{\partN \to 0} \ZZ_\partN^{\bv} (t) = 1$.
We can therefore focus on $\lambda_{\partM} \equiv \lim_{\partN \to 0} \partN \, \QQ_{\partN, \partM}$.
Although in principle $\partN \to 0$ would imply a vanishing number of variables,
$\lambda_{m}$ is well-defined as a symmetric polynomial and  
admits an explicit expansion in terms of the elementary symmetric 
polynomials $e_{p}$ in the ring of symmetric functions. 

In a similar way, we define $\tilde{\lambda}_{\partM} (\pv)$ from the limit of $n \, \tilde{\QQ}_{n,m} (\pv)$. 
The latter can be expressed by taking the limit $\partN \to 0$ of \eqref{LambdatildeFin}
\begin{equation}
\label{lambdan0}
\tilde{\lambda}_{\partM} =  
\sum_{p=0}^{2\partM} \frac{(-1)^{p-1} m! (m-1)! e_{2\partM - p}e_{p}}{(p-1)! (2\partM - p -1)! } 
= - \partM! (\partM-1)! [ H_0(x) H_0(-x)]_{x^{2m}}\;, \qquad  
H_0(x) = \sum_{p=0}^\infty \frac{e_p x^p}{(p-1)!}  \;.
\end{equation}
using that $(n-p)! \simeq (-1)^{p-1}/(n (p-1)!)$ for strictly positive integer $p$ and small $\partN$ and 
introducing the auxiliary function $H_0(x)$.
Note that in the expansion of $\tilde\lambda_{\partM}$ both the terms $p = 0$ and $p=2\partM$ give
a vanishing contribution.

We can now easily express the $\lambda_{m}$. Since 
\begin{equation}
\Omega_{\partM}^a \equiv \lim_{\partN\to0} \Omega_{\partN, \partM}^a = 
\frac{m!(m-1)! (-1)^a B_{2 a}^{(2 m-1)}(m)}{(m-a)! (m-a-1)! (2 a)!}
\end{equation}
is not singular,
we can take the limit directly in \eqref{recurstilde}, and arrive at
\begin{equation}
\label{recursn0}
 \lambda_{\partM} (\mu) = 
 \sum_{a=0}^{\partM} c^{2a} \Omega_{\partM}^a \tilde\lambda_{\partM-a} (\mu) 
 =- \partM! (\partM-1)! [ \omega_{0,\partM}(i c x) H_0(x) H_0(-x)]_{x^{2m}}
 \;.
\end{equation}
where 
\bea
&&  \omega_{0,m}(x) = \lim_{\partN \to 0}  \omega_{n,m}(x) = 
 \frac{1}{m! (m-1)!}\sum_{a=0}^m (-1)^a \Omega_{\partM}^a 
 (m-a)! (m-a-1)!  x^{2a} \nonumber \\
 && = \frac{1}{2} (G_{2m-1}(x,m) + G_{2m-1}(-x,m)) + O(x^{2m + 2})\label{omegalim1}
\eea
in agreement with (\ref{soluomega}), as expected. 
Again, as above in Eq.~\eqref{soluomega}, the orders higher than $x^{2m}$ are irrelevant for the final result.

\subsection{Large time limit}
We now turn to the calculation of $\Theta_{\partN, \partM}$. 
First of all, we need to replace, in the conserved charges, the values of the rapidities with a ``string state''.
Since the charges $\AA_p(\pv)$ are additive we have $\AA^s_p(\kv, \mv) = \sum_{j=1}^{n_s} \aA_p(k_j, m_j)$, where 
$A_p(k_j, m_j)$ are the contributions relative to a single string. As shown in Appendix \ref{app_charges}, 
they can be written as
\begin{equation}
\label{chargeonstring}
 \aA_p(k, m) = \sum_{q=0}^p \binom{p}{q} (i c)^{p-q} (2^{q-p+1}-1) B_{p-q} 
 \hA_q(k, m)	
\end{equation}
where $B_p = B_p^{(1)}(0)$ are the standard Bernoulli numbers and the homogeneous conserved charges, 
satisfying $\hA_p(u k, u m) = u^{p+1} \hA_p(k,m)$,
have been defined as
\begin{equation}
\label{homocharges}
 \hA_p(k, m)= 
 \frac{\left(k+\frac{i c
   m}{2}\right)^{p+1} - \left(k-\frac{i c m}{2}\right)^{p+1}}{i c (p+1)} \;. 
\end{equation}
As argued in \cite{de2015crossing}, the string average of products of homogeneous charges has
a simple scaling with $t$ at large times:
\begin{equation}
 \label{timescaling}
\lim_{\partN \to 0} \frac {\langle \hA_{p_1} \ldots \hA_{p_r}\rangle_{\partN}}{n} \propto_{t\to\infty} t^{[1 - (p_1+1) - \ldots - (p_r+1)]/3} \;.
\end{equation}
Since $\aA_1(k,m) = \hA_1(k,m)$ , 
the leading contribution $O(t^{-1})$ to each moment $\overline{p^m}$ 
is then given by the terms involving 
$\langle (\AA_1)^2 \rangle_{\partN} = \partN/(2 t)$ as dictated by STS
(for detailed version of these arguments and this last identity see \cite{supplmat_deluca2015}). 
Then, when expanding $e_p$ as a function of the $\AA_p$ through \eqref{newt},
we do not need the higher charges, $\AA_{p>1}$, as they will give subleading contributions to the moments at large time.
Combining \eqref{newt} and \eqref{chargeonstring}, we observe 
\begin{equation}
 e_p^s(\kv, \mv) = \frac{\AA_p^s (\kv, \mv)}{p} + \mbox{``terms involving more than one $\AA$''} 
 \longrightarrow (i c)^{p-1} (2^{2-p}-1) B_{p-1} \AA_1^s(\kv, \mv) \;.
\end{equation}
where in order to derive the last replacement we used \eqref{chargeonstring} and $\hA_1(k,m) = \aA_1(k,m)$.
Once this replacement is applied in Eq.~\eqref{lambdan0}, it leads to 
\begin{equation}
H_0(x| \pv^s) \longrightarrow x G_1 \bigl(i \cbar x,\frac{1}{2}\bigr) \AA_1^s(\kv, \mv) \;.
\end{equation}
This function is now odd: $H_0(-x| \pv^s) = H_0(x|\pv^s)$, as expected since $\hA_1$ only appears
in the expansion $\aA_p$ for odd $p$'s.
Finally using Eq.~\eqref{recursn0} and Eq.~\eqref{omegalim1}, 
we obtain the moments at large times
\begin{align}
 \label{moment}
 \overline{p^m} \simeq \lim_{\partN \to 0} \frac{\langle \lambda_{m} \rangle_\partN}{\partN} &=
 \lim_{n\to 0} m!(m-1)! \frac{\langle [\omega_{0,m}(i \cbar x) H_0(x)^2]_{x^{2m}}\rangle_n}{n}
 \longrightarrow \frac{m!(m-1)!}{2t} [x^2 G_{2 m+1}(i \cbar x,m+1)]_{x^{2m}} \\ 
 &= 
 \frac{(-1)^{m-1} \cbar^{2m-2} (m-1)! m! B_{2 m-2}^{(2 m+1)}(m+1)}{2 (2 m-2)! t}
 = \frac{(m-1)!^4 \cbar^{2m-2} }{2 (2 m-1)! t} \;.
\end{align}
where in the first line we have used the multiplication formula for the Bernoulli generating functions 
(see end of Appendix \ref{app_omega} ) and in the second line we have
used the value (\ref{res2}) of the Bernoulli polynomial at a special argument, obtained in the Appendix \ref{app_genbern}.  

Thus, we have shown as announced in the introduction, that, for integer $m \geq 1$
\begin{equation}
\label{finalresult}
\overline{p^m} =_{t \to +\infty} \frac{\gamma_m \cbar^{2m-2}}{t} + o(t^{-1}) \;, \qquad \gamma_m =  
 \frac{\sqrt{\pi } 4^{-m} \Gamma (m)^3 }{\Gamma \left(m+\frac{1}{2}\right) } 
\end{equation}
where we have rearranged the Gamma functions in the $\gamma_m$. It is easy to check that
the values for $m=1,2,3$ given in Eq.~\eqref{lowmom} are recovered.

\subsection{Final result and comparison with numerics \label{sec_comparison}}
We want now to recover the density $\rho(p)$ associated to the moments in Eq.~\eqref{finalresult}.
For simplicity, in this section we set $\cbar = 1$. The full result can be recovered by rescaling
as in \eqref{conj2}. We look for a function $\rho(p)$ satisfying
\begin{equation}
 \int_0^\infty d\rho \, \rho(p) p^m = \gamma_\partM \;.
\end{equation}
for all integers $m \geq 1$.
One can note that the densities $\rho_1 (u) = e^{-u}/u$ and $\rho_2(u) = \theta(0<u<1) (1-u)^{-1/2}/u $
have respectively moments $\overline{u^m}=\Gamma(m)$ and 
$\overline{u^m}=\sqrt{\pi} \Gamma(m)/\Gamma(m+1/2)$.
We then obtain, by convolution,
the density
\begin{equation}
\label{rho1}
 \rho(p)= \frac{2}{p} \int_0^{+\infty} \frac{du}{\sqrt{ u (u+4)}} K_0(2 \sqrt{p} \sqrt{u+ 4})
\end{equation}
An equivalent expression, suited for asymptotic expansion at small $p$, is given by the contour integral
\bea
\label{rho2}
\rho(p) = \frac{1}{p} \int_{\epsilon - i \infty}^{\epsilon + i \infty} \frac{ds}{2 i \pi} p^{-s} 4^{-s} \sqrt{\pi} \Gamma(s)^3/\Gamma(\frac{1}{2}+s)
\eea 
for a small $\epsilon>0$. 
For $p<1$, the contour can be closed on the
half plane $\Re(s) < 0$ and one gets the sum of residues as an expansion in small $p$: the
first term (residue in zero) gives $ (\ln p + 2 \gamma_E)^2/(2 p)$, the 
second one gives $p^2$ and $p^2 (\ln p)^2$ and so on. 
As explained in the introduction, this suggests that the validity of this tail for large $p$, extends
up to a cut-off value $p_c$ bounded by $p_c \gg e^{-1.817 t^{1/3}}$.
\\

We now compare our analytical result for the continuum model 
with the discrete directed polymer on a square lattice \cite{calabrese2010free},
defined according to the recursion (with integer time $\hat t$ running along the diagonal) 
\begin{equation}
 Z_{\hat x, \hat t + 1} =  (Z_{\hat x - \frac{1}{2},t} + Z_{\hat x + \frac{1}{2}, \hat t} ) e^{-\beta V_{\hat x, \hat t + 1}}
\end{equation}
with $V_{\hat x, \hat t}$ sampled from the standard 
normal distribution. 
This discrete model reproduces the continuous DP 
in the high temperature limit $\beta \ll 1$,
under the rescalings: $\cbar = 1$ with $x = 4\hat x \beta^2$ and $t = 2\hat t \beta^4$ 
\cite{calabrese2010free}. 
As done in \cite{de2015crossing}, we take
two polymers with initial conditions $Z_{\hat x, \hat t = 1}^{\pm} = \delta_{\hat x, \pm 1/2}$
and ending at time $\hat t$ at 
$\hat x = \pm 1/2$. Then,
for each realization, the non-crossing probability $\hat p$ on the lattice 
is computed by the image method 
\cite{karlin1959coincidence, *gessel1985binomial, *gessel1989determinants, *wiki:lgvlem}.
The relation between $\hat p$ on the lattice and the random variable $p$ can be read from
\eqref{plimit}, which leads to $\hat p \simeq 16 p \beta^4$, for $\beta \to 0$.
As shown on the figure \ref{comparison} the agreement between the
numerics, in the double limit $\hat t\to \infty$ and $\beta \to 0$, and 
our prediction for $\rho(p)$ is convincing.

\section{Conclusion}
We presented an exact method to compute the large-time asymptotics of 
the moments of the non-crossing probability for two polymers in a random medium.
As an intermediate outcome, an algebraic approach, based on generating functions, is developed 
to express explicitly a class of symmetric polynomials,
related to arbitrary number of replicas of two mutually-avoiding polymers.
In the large-time limit, the calculation of the moments further simplifies 
and an analytic expression is provided. 
In this way, an explicit formula, compatible with these moments, for the tail of the full distribution of the
non-crossing probability is proposed. Its validity is then benchmarked against
numerical simulations on a discretization of the continuous directed polymer problem. 

This approach provides a rare analytical result in the complicated interplay between disorder and interactions.
Moreover, several new perspectives and generalizations become accessible. First of all, a larger number 
of mutually avoiding polymers is treatable within the same framework. Then,  
the next question, currently under investigation by the authors,
concerns the bulk of the distribution. The conjectured 
connection with the statistics of the first few eigenvalues 
of a random Gaussian matrix should be addressable within our approach.

\paragraph{Acknowledgments. ---}
We thank A. Borodin, I. Corwin, and A. Rosso for interesting discussions. 
This work is supported by ``Investissements d'Avenir'' 
LabEx PALM (ANR-10-LABX-0039-PALM)
and by PSL grant ANR-10-IDEX-0001-02-PSL.

\appendix
\section{Explicit formula for $\tilde\Lambda_{n,m}$ \label{app_tildel}}
We now show that, for given $\partN, \partM$, any polynomial $\tilde{\QQ}_{\partN,\partM}(\pv)$ satisfying
properties 1-4 presented in Sec. \ref{czerocase} equals, up to a multiple, the expression in Eq.~\eqref{LambdatildeFin}.
Clearly, being symmetric, it admits a representation
in terms of elementary symmetric polynomials. 
Moreover property 3 implies that it is a quadratic function of the $e_p$'s 
and from homogeneity we arrive at
\begin{equation}
 \label{tildehomo}
 \tilde{\QQ}_{\partN, \partM}(\pv) = \sum_{p=0}^{2m} a_p e_p e_{2m - p} \;.
\end{equation}
with the coefficients $a_p$ satisfying $a_p = a_{2m-p}$. Using that 
\begin{equation}
e_p(\pv + a) = e_p(\pv) + (\partN - p + 1)\,a\, e_{p-1}(\pv) + O(a^2)\;,
\end{equation}
property 4 leads to
\begin{equation}
 \left.\frac{d \tilde{\QQ}_{\partN, \partM}(\pv + a)}{da}\right|_{a=0} = 
 \sum_{p=1}^{2m} [ a_p (n-p-1) + a_{p-1} (\partN - 2m + p)
]e_{p-1} e_{2m - p}  = 0\;.
\end{equation}
For this condition to be true for arbitrary values of $\pv$, we arrive at
\begin{equation}
 a_p = -a_{p-1} \times \left(\frac{\partN - 2m + p}{\partN-p-1}\right) =
 (-1)^{p-\partM} \frac{(\partN - p)! (\partN - 2 \partM + p)!}{[(\partN - \partM)!]^2} a_{\partM}
\end{equation}
Then, simple inspection of Eq.~\eqref{freeBoro} gives $a_m = \partM!(\partN-\partM)!/\partN!$ and
Eq.~\eqref{LambdatildeFin} follows.

\section{Characterization of $\Lambda_{\partN,\partM}$ \label{app_tildefull}}
\subsection{Polynomial from symmetrization\label{app_residue}}
In this subsection we show that $\QQ_{\partN, \partM}(\pv)$ defined in Eq.~\eqref{borodinG}
is actually a symmetric polynomial in the rapidities. 
More generally, we show that for any polynomial $q(\pv)$,
the rational function $\phi(\pv)$ defined by
\begin{equation}
\label{Phiproof}
\Phi(\pv) = \sym_{\pv} \left[ \frac{q(\pv)}{\prod_{\alpha<\beta} (\mu_\alpha - \mu_\beta) }\right]
\end{equation}
is itself a polynomial. Indeed, we can rewrite it as
\begin{equation}
\label{Phiproof1}
\Phi(\pv) = \frac{1}{\partN!\prod_{\alpha<\beta} (\mu_\alpha - \mu_\beta) }
\sum_{P \in \mathcal{S}_\partN} (-1)^{\sigma_P} q(\mu_{P_1}, \ldots, \mu_{P_n})
= \frac{\asym_{\pv}[ q(\pv) ]}{\prod_{\alpha<\beta} (\mu_\alpha - \mu_\beta) }
\end{equation}
where the anti-symmetrization operator $\asym[\ldots]$ has been introduced. Since $\asym[q(\pv)]$ is an alternating polynomial,
it will be a multiple of the denominator and therefore $\Phi(\pv)$ is itself a polynomial.

\subsection{Expansions of $\Lambda_{n,m}$ in powers of $c$ \label{app_taylorc}}
We show here that $\QQ_{\partN, \partM}(\pv)$ admits the expansion \eqref{recurstilde}.
First of all we notice that reverting the order of rapidities $\mu_\alpha \to \mu_{\partN - \alpha}$ 
is equivalent to sending $c \to - c$. Then, after symmetrization, $\QQ_{\partN, \partM}(\pv)$
will be an even function of $c$ and we can expand it as
\begin{equation}
\label{evenexp}
 \QQ_{\partN, \partM}(\pv) = \sum_{a=0}^{m} \cbar^{2a} P_{n,m,a}(\pv)
\end{equation}
where $P_{\partN,\partM,a}(\pv)$ are homogeneous and symmetric polynomial of degree $2\partM - 2a$.
As explained in Appendix \ref{app_tildel}, in order to proof that 
$P_{\partN, \partM,a}(\pv) \propto \tilde{\QQ}_{\partN, \partM-a}(\pv)$, we simply need to show
that $P_{\partN, \partM,a}(\pv) $ satisfies properties 1-4 of Sec. \ref{czerocase}. The only non-trivial
property is 3. But we can write
\begin{equation}
 \label{lambdader}
 P_{\partN, \partM, a}(\pv) = \frac{1}{(2a)!} \left.\frac{d^{2a} \QQ_{\partN, \partM}(\pv)}{d \cbar^{2a}}\right|_{c=0}
\end{equation}
and after applying all the derivatives, we obtain several terms of the form
\begin{equation}
\label{termderiv}
\sym_{\pv} \left[\frac{(\mu_{\alpha_1} - \mu_{\beta_1}) \ldots (\mu_{\alpha_{p + 2m - 2a}} - \mu_{\beta_{p + 2m - 2a}}) }{(\mu_{\gamma_1} - \mu_{\delta_1}) \ldots (\mu_{\gamma_{p}} - \mu_{\delta_{p}}) } \right]
\end{equation}
with $p = 0, \ldots, 2a$. As the numerator comes from $2a-p$ differentiations, 
with respect to $c$, of $\prod_{q=1}^m h(\mu_{2q-1,2q})$, each variables cannot appear more than twice. 
After symmetrization in Eq.~\eqref{termderiv} we obtain a polynomial, 
as explained in Sec.~\ref{app_residue}, and therefore each term satisfies property 3.

\subsection{Value on strings \label{app_strings}}
We show in this subsection that $\QQ_{\partN, \partM}(\pv)$
vanishes whenever the set of rapidities $\pv$ contains a $\ell$-string with $\ell > \partN - \partM$. 
To fix the notation we slightly extend Eq.~\eqref{stringlzero} to
\begin{equation}
\label{stringl}
 \pv^{\ell} = \left( \mu_1 = \frac{i \cbar }{2} (\ell - 1), \mu_2 = \frac{i \cbar }{2} (\ell - 3), \ldots,  
\mu_\ell = -\frac{i \cbar }{2} (\ell - 1), \mu_{\ell+1}, \ldots, \mu_{\partN} \right) \;.
\end{equation}
which reduces to $\pv^{\ell, 0}$ when $\mu_{\alpha} = 0$ for $\alpha>\ell$.
Note that the momentum of the $\ell$-string can be set to zero, without losing generality, as $\QQ_{\partN, \partM}(\pv)$
only depends on the differences between pairs of rapidities and the $\mu_{\alpha}$'s with $\alpha> \ell$  in Eq.~\eqref{stringl} are arbitrary.
Writing 
explicitly Eq.~\eqref{borodinG} and exchanging $c\to \cbar$ (it is an even function of $\cbar$ as showed in \eqref{evenexp}), we have 
\begin{equation}
 \label{borodinGexpl}
 \QQ_{\partN,\partM}(\pv) = \sum_{P \in \mathcal{S}_\partN} \left[\prod_{q=1}^m (\mu_{P_{2q -1}} - \mu_{P_{2q}})(\mu_{P_{2q -1}} - \mu_{P_{2q}} - i\cbar)\right] 
 \left[\prod_{\alpha<\beta} \frac{\mu_{P_\beta} - \mu_{P_\alpha} - i \cbar}{\mu_{P_\beta} - \mu_{P_\alpha}}\right]
\end{equation}
and it is clear that the numerator of the second product will vanish unless the order of the first
$\ell$ rapidities is left unchanged by the permutation $P$:
$P^{-1}_{\alpha+1} > P^{-1}_{\alpha}$ for all $\alpha = 1,\ldots, \ell-1$. Instead,
the first product will vanish whenever $P_\alpha^{-1} = 2q-1, P_{\alpha+1}^{-1} = 2q$ 
for some $q = 1,\ldots, m$ and $i = 1,\ldots, \ell-1$.
These two conditions are compatible only for $\ell \leq \partN - \partM$. In particular, 
in the limiting case $\ell = \partN - \partM$, only two types of permutations are possible:
\begin{equation}
  \left(\mu_{P_1}, \mu_{P_2}, \ldots ,\mu_{P_\partN}\right) \\
 = \left\{\begin{array}{l}
\left(  \mu_{1}, x, \mu_{2}, x, \mu_{3}, \ldots, \mu_{\partM}, \mu_{\partM+1}, \ldots, \mu_{\partN - \partM}\right) \\
\left(  x, \mu_{1}, x, \mu_{2}, x,  \ldots, x, \mu_{\partM}, \mu_{\partM+1}, \ldots, \mu_{\partN - \partM}\right)
 \end{array}\right.
\end{equation}
where the x's stand for arbitrary permutations of the remaining $\partM$ rapidities. Then, 
it is clear that for $\ell > \partN - \partM$ at least two consecutive rapidities of 
the $\ell$-string would be adjacent in the first $2\partM$ places, 
and all the terms in the sum \eqref{borodinGexpl} for arbitrary $P$ would vanish.

\subsection{Calculation of the coefficients $\Omega_{\partN, \partM}^a$ \label{app_omega}}
As shown in Eq.~\eqref{recursGen} of the text, $\QQ_{\partN, \partM} (\pv)$ 
can be written employing generating functions and the function $\omega_{\partN, \partM}(x)$
contains all the unknowns. We use conditions in \eqref{substl} to fix the function $\omega_{\partN,\partM}(x)$.
First we note that 
\begin{equation}
 E(x | \pv^{\ell,0}) = 
 \frac{(- i \cbar x)^\ell \Gamma \left( \frac{1 + \ell}{2} + \frac{i}{\cbar x}\right)}{\Gamma \left( \frac{1-\ell}{2} + \frac{i}{\cbar x}\right)}
 \;.
\end{equation}
Then using the asymptotic expansion \cite{tricomi1951asymptotic} for $z\to \infty$
\begin{equation}
\label{largegamma}
 \frac{\Gamma(z + \alpha)}{\Gamma(z + \beta)} = z^{\alpha-\beta} \sum_{n=0}^\infty (-1)^n \frac{(\beta-\alpha)_n}{n!} B_n^{(\alpha - \beta + 1)}(\alpha) z^{-n}
\end{equation}
and $z=i/(\cbar x)$, we deduce \eqref{symstring}.
Then injecting in Eq.~\eqref{Hfun}
\begin{equation}
 \label{Hfun1}
 H_\partN(x | \pv^{\ell,0}) = \sum_{p=0}^\partN (\partN-p)! e_p(\pv^{\ell,0}) x^p = 
 \ell! \sum_{p=0}^\partN \frac{(\partN-p)!}{(\ell - p)!}  \frac{B_p^{(\ell+1)}\left(\frac{\ell+1}{2}\right) (- i \cbar x)^p }{p!}   \;.
\end{equation}
One easily sees from its definition that $B_p^{(\ell+1)}\left(\frac{\ell+1}{2}\right)=0$ for $p$ odd, which implies
that the function $H_\partN(x | \pv^{\ell,0})$ is even in $x$. 
In this last sum, we can safely replace the upper bound for $p$ to $+ \infty$, 
since higher powers in $x$ will not affect
$\QQ_{\partN,\partM}(\pv)$ in \eqref{recursGen}. Then, from the definition in Eq.~\eqref{genbern}, we obtain
for $\ell=n$:
\begin{equation}
 \label{Hfunn}
 H_\partN(x | \pv^{\partN,0}) = n! G_{n+1} \left(- i \cbar x, \frac{n+1}{2}\right) + O(x^{n + 1})
\end{equation}
together with the recursive relation in $\partN$
\begin{equation}
 \label{HfunRec}
 H_{\partN+1}(x | \pv^{\ell,0}) = - x^{\partN + 2} \frac{d}{dx}[x^{-n - 1} H_{\partN}(x | \pv^{\ell})] \;.
\end{equation}
Then, using the relation
\begin{equation}
 x^{\partN+2} \frac{d}{dx}\left[ x^{-\partN-1} G_{\partN +1} (x,y)\right] = (y - \partN - 1) G_{\partN+2} (x, y+1)
 - y G_{\partN + 2}(x,y)
\end{equation}
it is easy to prove, by induction over $\partN$, starting from $\partN = \ell$, that
\begin{equation}
 H_{\partN}(x | \pv^{\ell,0}) = \sum_{p=1}^{\partN - \ell +1} a_p G_{n+1}\bigl(-i \cbar x, \frac{\ell -1 + 2p}{2}\bigr) + O(x^{n+1})\;.
\end{equation}
for appropriate coefficients $a_p$ which explicit values are not needed below. Then, taking the square of this expression, the multiplication
formula $G_\alpha(x,y) G_\beta(x,z) = G_{\alpha + \beta} (x, y + z) $ leads to Eq.~\eqref{HH}.

\subsection{Special values of generalized Bernoulli polynomials}
\label{app_genbern} 

Fixing an integer $p$, one has
\bea
B_{p} ^{(p+1)}(y) = p! [ (\frac{x}{e^x-1})^{p+1} e^{x y} ]_{x^{p} }
= p! \int _C \frac{dz}{2 i \pi z} z^{-p}  (\frac{z}{e^z-1})^{p+1} e^{z y}
\eea 
where $C$ is a small contour around the origin. This simplifies
into
\bea
B_{p} ^{(p+1)}(y) = p!  \int _C \frac{dz}{2 i \pi} (e^z-1)^{-p-1} e^{z y} 
=  p!  \int _C \frac{dw}{2 i \pi} \frac{(1+w)^{y-1}}{w^{p+1}}
= (y-p)_{p} 
\eea 
where we have changed $e^z-1=w$. More generally, for integer $q$
\begin{equation}
B_{p-q} ^{(p+ 1)}(y) = (p-q)! [ (\frac{x}{e^x-1})^{p+1} e^{x y} ]_{x^{p-q} }
= (p-q)! \int _C \frac{dz}{2 i \pi} z^{q}  (e^z-1)^{-p-1} e^{z y}
= \frac{(p-q)!}{p!}\frac{d^q}{dy^q} (y-p)_p \;.
\end{equation}
It follows for $p = 2m$ and $q = 2$
\begin{align}
\label{res2}
&B_{2 m-2}^{(2 m+1)}(m+1) = \frac{1}{2m(2m-1)}\left.\frac{d^2}{dy^2} (y-2m)_{2m}\right|_{y = m+1} = 
\frac{(-1)^{m - 1} (m - 1)!^2}{m(2m-1)}\\
\end{align}

\section{Conserved charges on strings \label{app_charges}}
The value of the conserved charges on a single string is defined as
\begin{equation}
 \label{singlestringcharge}
 A_p(k, m) = \sum_{a=0}^{m-1} \left( k + \frac{i \cbar (m-1 - 2 a)}{2}\right)^p \;.
\end{equation}
In order to compute this sum, we introduce the charge exponential generating function
\begin{equation}
\label{Agen}
\mathfrak{A}(x) = \sum_{p=0}^\infty \frac{A_p x^p}{p!} = e^{k x} \sum_{a=0}^{m-1} \exp{ \left(\frac{i \cbar x (m-1 - 2 a)}{2}\right)}
=  \frac{2 e^{k x}\sin (\frac{m \cbar x}{2})}{\cbar x} G_1\bigl(i \cbar x, \frac{1}{2}\bigl)
\end{equation}
using the definition (\ref{genbern}) of $G_1$. 
From this expression, it is clear that the denominator present in $G_1$  
produces the ``inhomogeneity'' in the expansion of $A_p(k, m)$. 
Therefore, if we 
define the generating function of the homogeneous charges as 
\begin{equation}
 \mathfrak{A}^{(h)}(x) \equiv \frac{2 e^{k x}\sin (\frac{m \cbar x}{2})}{ \cbar x} = \sum_{p=0}^\infty \frac{\hA_p(k,m) x^p}{p!}
\end{equation}
we immediately deduce Eq.~\eqref{homocharges}. Then Eq.~\eqref{chargeonstring} follows combining
Eq.~\eqref{Agen} and Eq.~\eqref{genbern} and using 
that $G_1(x, \frac{1}{2}) = 2 G_1(\frac{x}{2},0)-1$.

\AtEndEnvironment{thebibliography}{
\bibitem{Doumerc}
N. O'Connell, M. Yor, Elect. Comm. in Probab. 7 (2002) 1,
Y. Doumerc, Lecture Notes in Math., 1832: 370 (2003),
I. Corwin et al., arXiv:1110.3489v4.

\bibitem{SoOr07}
A. M. Somoza, M. Ortu\~no and J. Prior, 
Phys. Rev. Lett.  \textbf{99}, 116602 (2007). A. Gangopadhyay, V. Galitski, M. Mueller, arXiv:1210.3726,
Phys. Rev. Lett. 111, 026801 (2013).
A. M. Somoza, P. Le Doussal, M. Ortuno, 
arXiv:1501.03612 (2015). 
\bibitem{bec}
J. Bec, K. Khanin, arXiv:0704.1611, Phys. Rep. 447, 1-66, (2007). 
%

\bibitem{png}M. Pr\"ahofer and H. Spohn,
Phys. Rev. Lett. \textbf{84}, 4882 (2000);
J. Baik and E. M. Rains, J. Stat. Phys. \textbf{100}, 523 (2000).
 \bibitem{spohnKPZEdge}
T. Sasamoto and H. Spohn, Phys. Rev. Lett. {\bf 104}, 230602 (2010);
Nucl. Phys. B {\bf 834}, 523 (2010); J. Stat. Phys. {\bf 140}, 209 (2010).
\bibitem{corwinDP}
G. Amir, I. Corwin, J. Quastel, Comm. Pure Appl. Math {\bf 64}, 466 (2011).
I. Corwin, arXiv:1106.1596. 
\bibitem{calabreseSine}
P. Calabrese, M. Kormos and P. Le Doussal, EPL {\bf 107}, 10011 (2014)
\bibitem{Quastelflat} 
J. Ortmann, J. Quastel and D. Remenik 
arXiv:1407.8484
and arXiv:1501.05626.
%
\bibitem{dotsenko}
V. Dotsenko, EPL {\bf 90}, 20003 (2010); J. Stat. Mech. P07010 (2010);  
%
\bibitem{flat}
P. Calabrese and P. Le Doussal, Phys. Rev. Lett. {\bf 106}, 250603 (2011)
and J. Stat. Mech. (2012) P06001.
\bibitem{SasamotoStationary}
T. Imamura, T. Sasamoto, arXiv:1111.4634, 
Phys. Rev. Lett. {\bf 108}, 190603 (2012); arXiv:1105.4659,
J. Phys. A  {\bf 44},  385001 (2011); and arXiv:1210.4278
J. Stat. Phys. 150, 908-939 (2013).
\bibitem{natter}
T. Nattermann, I. Lyuksyutov and M. Schwartz, EPL 16 295
(1991), J. Toner and D.P. DiVicenzo, Phys. Rev. B 41, 632(1990).
J. Kierfeld and T. Hwa, Phys. Rev. Lett. 77, 20, 4233 (1996). 

\bibitem{tricomi1951asymptotic}
Tricomi, F. G. and  Erdélyi, A., Pacific J. Math, 1(1), 133-142 (1951).

\bibitem{ll} E. H. Lieb and W. Liniger, Phys. Rev. {\bf 130}, 1605 (1963).

\bibitem{Endpoint}
V. Dotsenko, arXiv:1209.6166, G. Schehr, arXiv:1203.1658,
G. R. Moreno Flores, J. Quastel, and D. Remenik, arXiv:1106.2716; 
J. Quastel and D. Remenik,  arXiv:1111.2565, 
J. Baik, K. Liechty, G. Schehr, arXiv:1205.3665, J. Math. Phys. 53, 083303 (2012).

\bibitem{2point}
S. Prolhac and H. Spohn, arXiv:1011.401, J. Stat. Mech. (2011) P01031,
S. Prolhac and H. Spohn, arXiv:1101.4622, J. Stat. Mech. (2011) P03020,
V. Dotsenko, arXiv:1304.6571,
T. Imamura, T. Sasamoto, H. Spohn, arXiv:1305.1217, I. Corwin and J. Quastel, arXiv:1103.3422.

\bibitem{Pimentel}
L. P. R. Pimentel, arXiv:1207.4469. 
S. I. Lopez and L. P. R. Pimentel, arXiv:1510.01552

\bibitem{markus_magneto} 
A. Gangopadhyay, V. Galitski, M. Mueller,
arXiv:1210.3726 
Phys. Rev. Lett. 111, 026801 (2013)

\bibitem{supplmat_deluca2015}
Supplemental Material of \cite{de2015crossing}
at \url{http://link.aps.org/supplemental/
10.1103/PhysRevE.92.040102}.
}

\bibliography{KPZ}

\end{document}